\documentclass[runningheads]{llncs}

\usepackage{wrapfig}
\usepackage{multirow}
\usepackage{graphicx}
\usepackage[table,xcdraw]{xcolor}
\usepackage{tabularx}
\usepackage{graphicx}
\usepackage{adjustbox}
\usepackage{tikz}

\usepackage[T1]{fontenc}
 \usepackage{multirow}
 \usepackage{pdflscape}
\usepackage[table,xcdraw]{xcolor}
\usepackage{graphicx}
\usepackage{color}

\usepackage{graphicx}
\usepackage{wrapfig}
\usepackage{lscape}
\usepackage{rotating}
\usepackage{epstopdf}
\usepackage{cite}

\usetikzlibrary{positioning,shadows}

\tikzset{mynode/.style={draw=white,solid,circle,fill=green,inner sep=1pt, thick,
text=black}}

\newcommand*\samethanks[1][\value{footnote}]{\footnotemark[#1]}
\begin{document}
\title{An Empirical Study on How the Developers Discussed about Pandas Topics}
%
%
\author{Sajib Kumar Saha Joy\thanks{First Three authors contributed equally to this research.}\and
Farzad Ahmed\samethanks \and
Al Hasib Mahamud\samethanks \and
Nibir Chandra Mandal}
\authorrunning{Joy et al.}
%
\institute{Ahsanullah University of Science and Technology, Bangladesh\\
\email{ \{joyjft, farzadahmed6\}@gmail.com, \{hasib, nibir\}.cse@aust.edu}}
%
\maketitle              

\begin{abstract}
Pandas is defined as a software library which is used for data analysis in Python programming language. As pandas is a fast, easy and open source data analysis tool, it is rapidly used in different software engineering projects like software development, machine learning, computer vision, natural language processing, robotics, and others. So a huge interests are shown in software developers regarding pandas and a huge number of discussions are now becoming dominant in online developer forums, like Stack Overflow (SO). Such discussions can help to understand the popularity of pandas library and also can help to understand the importance, prevalence, difficulties of pandas topics. The main aim of this research paper is to find the popularity and difficulty of pandas topics. For this regard, SO posts are collected which are related to pandas topic discussions. Topic modeling are done on the textual contents of the posts. We found 26 topics which we further categorized into 5 board categories. We observed that developers discuss variety of pandas topics in SO related to error and excepting handling, visualization, External support, dataframe, and optimization. In addition, a trend chart is generated according to the discussion of topics in a predefined time series. The finding of this paper can provide a path to help the developers, educators and learners. For example, beginner developers can learn most important topics in pandas which are essential for develop any model. Educators can understand the topics which seem hard to learners and can build different tutorials which can make that pandas topic understandable. From this empirical study it is possible to understand the preferences of developers in pandas topic by processing their SO posts.

\keywords{Pandas  \and Stack Overflow \and Natural Language Processing \and Empirical Software Engineering.}
\end{abstract}
\section{Introduction}
Pandas is an open source library where python package offers data manipulation and analysis in python programming language. As pandas has high performance and fast productivity for users, it's data analysis capability is utilized in different sectors of computer science like data visualization, machine learning, data driven software engineering, computer vision, natural language processing etc.  Since 2008, the development of pandas has removed the distance between the availability of data analysis tools \cite{ref_1}. Pandas is considered for most suitable option for data analysis tool as it is written in python programming language and easy to understand for new beginner \cite{ref_2}.\\
In the recent years, the utilization of pandas library is increased rapidly as pandas library has reduced the gap between scientific programming languages and database languages \cite{ref_99}. Pandas library is utilized in most of the sectors like machine learning, statistics, natural language processing, computer vision and others. Moreover, pandas library is easy to understand for a beginner and it is open source tool. For these reasons, most of the developers are now showing interest to utilize pandas tools in their projects. For the development of pandas library and it's utilization, a factor is observed that the discussions regarding pandas in online developers forums has increased, such as Stack Overflow (SO). From analysing these post, several findings can be achieved related to pandas library like it's popularity, difficulties, future scopes etc.  To date, there are around 22.44 million questions are posted in SO \cite{ref_url1}.Several research works are conducted based on SO posts in the field of IoT \cite{ref_3}, blockchain \cite{ref_4}, microservices \cite{ref_5}, software engineering\cite{ref_3}.Some research works are also done based on the functionality, popularity, scope to development of pandas library \cite{ref_7,ref_8,ref_9}. According to the best of our knowledge, no research work is done based on the SO posts of pandas library to find the topics, popularity, scopes of pandas library.\\
In this research paper, total 236711 SO posts where user defined tags are related to pandas are analyzed to find the topics of pandas library. For topic modeling, Latent Dirichlet Allocation (LDA) is performed. Finally trend chart is generated to find the popularity's of the topics according to the discussions of the software development forums. In this empirical study, some major findings are shown. Among of the major findings, firstly we have found the topics and then we have categories pandas topics which are discussed most in the SO posts. According to the findings of this paper, there are total twenty six topics and these twenty six topics can be categorized into six categories. Among of the topics optimization is the most popular topic though SQL queries and Matplotlib support are the most difficult topics as SQL queries is having the lowest score and Matplotlib support is having the lowest accepted answer rate. Secondly, to make a closer look of the topics and categories, a trend chart is generated from the time slot July 2011 to February 2022. Some decline and arises are seen in the trend chart of the topics but the total number of posts are increased gradually as the total amount of pandas developers increased over time.  
\\The next of the paper is organized in the following way: Section 2 discusses the background studies of this paper. Methodology is described in Section 3 where data collection, topic modeling and topic naming process are answered. Section 4 discusses implication of studies where several important expositions are described.  Section 5 describes threats to the validity of our result. Section 6 describes results of our study where section 7 answers the future scopes to work from the result. Section 8 concludes the paper.

\section{Background Studies}
\subsection{Stack Overflow}
Stack Overflow (SO) is considered as a question and answering sites which has become popular in recent times for software developers forum. There are some sites for programmers where programmers can ask questions, can answer other's questions and can share ideas. SO is one of them \cite{ref_11}.
\subsection{Topic Modeling}
Topic modeling can be defined as unsupervised classification method which is similar to clustering data and it is usually used for finding groups of item \cite{ref_url3}.Though topic modeling is mostly used for textual data, it can be used for bioinformatics data, social science data and other source of data \cite{ref_5}.\\
To apply topic modeling in the SO posts related to pandas, in this paper Latent Dirichlet Allocation (LDA) is used. In LDA, each document is considered as combination of topics and each topic is considered as combination of words \cite{ref_url3}. Topic modeling is used in different papers of various fields where SO posts are utilized, specially in the aspects of software engineering development \cite{ref_3}.
\subsection{Pandas}
Pandas is an open source data analysis tool which is built on Python programming language and it is considered as fast, flexible and easy tool \cite{ref_url3}. Pandas works in structured dataset and can be leveraged in social science, statistics, finance and other fields \cite{ref_99}. Since the development of pandas in 2008, the main aim of pandas library development is to remove the distance between scientific  programming languages and database languages \cite{ref_99}.
\subsection{Related Works}
There are several works related to LDA Topic Modeling \cite{ref_3,ref_15, ref_14,ref_12, ref_16,ref_3, ref_17, ref_18,ref_3}. 
Uddin et al. \cite{ref_3} have done an empirical study based on the IoT discussions of IoT topic on SO posts. For this purpose, authors have gathered IoT posts from SO and have leveraged LDA to perform topic modeling. Four research questions are answered to find the discussion, evolve, question types, popularity and difficlties of IoT topics.\\Stephen W. Thomas \cite{ref_15} has leveraged statistical topic modeling in mining software repositories to analyze unstructured and unlabeled data  and has found structure in the textual repositories. Bavota et al. \cite{ref_14}  have shown the opportunities of Move Method refactoring and have removed Feature Envy bad smells from source code and for this purpose Relational Topic Models are utilized.Nabil et al. \cite{ref_12} have utilized Topic modeling in cloud computing and to discover efficient cloud services, LDA is leveraged. To find the research trends, methodology and fields of further research in blockchain technology, Shahid et al. \cite{ref_16} have used topic modeling for literature analysis of the research. Mandal et al. \cite{ref_3} have utilized statistical topic modeling technique in software system and have found software concerns as topic. Ramage et al. \cite{ref_17} have utilized topic modeling in social sciences to find the barriers of adoption of topic modeling in social science field.  Mei et al. \cite{ref_18} have utilized topic modeling in network regularization to explain the difficulties of topic modeling with network structure.

\section{Methodology}
\begin{figure}
\centering
\hspace*{-2.75cm}
\includegraphics[scale = 0.4]{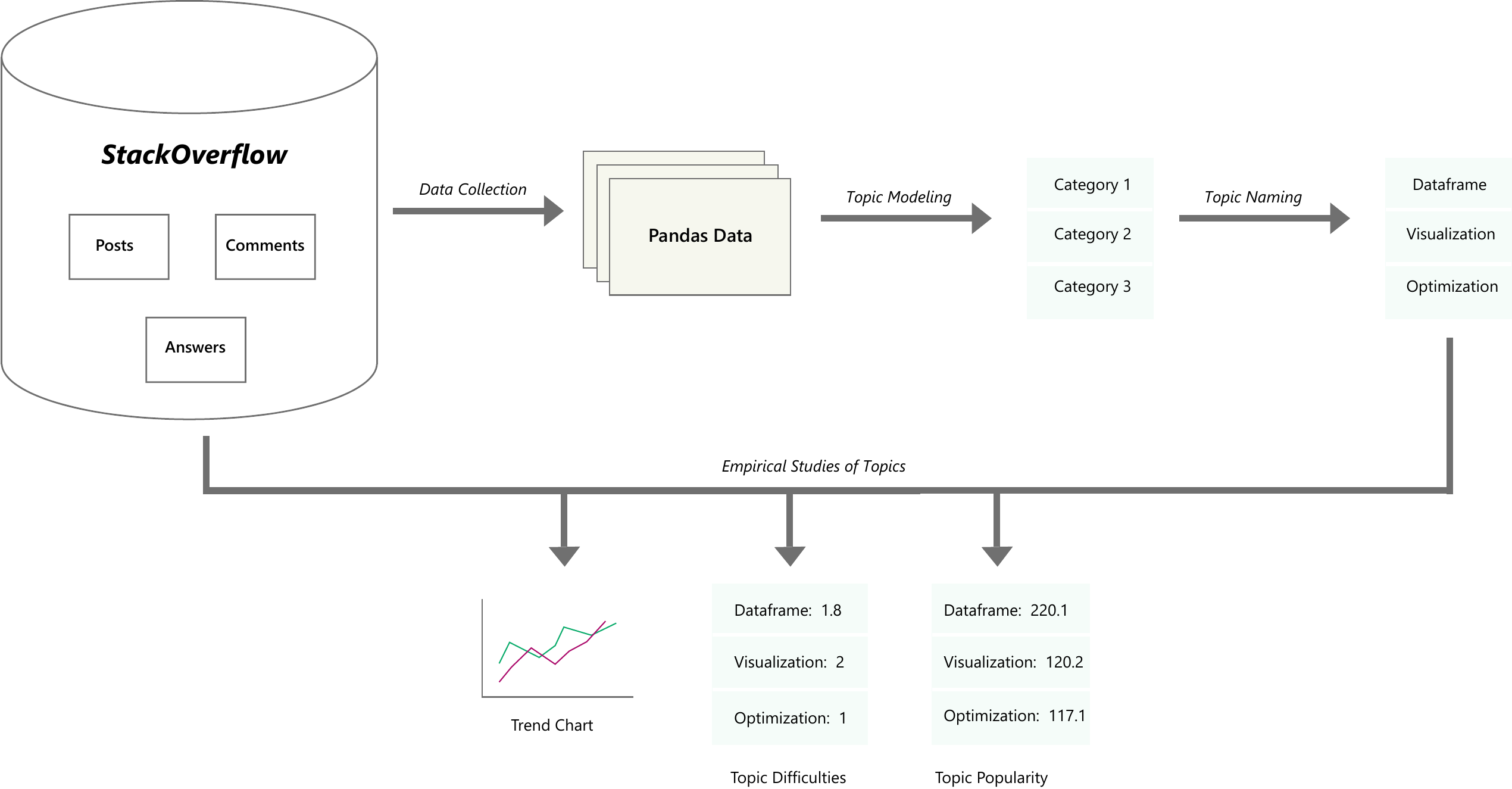}
\hspace*{-2.75mm}%
\caption{Proposed Methodology} \label{proposed_methodology}
\end{figure}

In this research work, we followed six steps to analyze pandas topic. As there has been no prior work, we began by collecting panda datasets from SO.We collected meta information such as post titles, creation dates, view counts, accepted answers, etc. for pandas posts from SO. Second, we applied LDA topic modeling to get all the discussed pandas topics in SO. From these steps, we only found categorized posts. However, LDA does not provide the topic name associated with each category. Thus, we followed the third step to name each category. In step four, we analyzed the topic using the collected meta information. We created a trend chart to understand how the pandas topic has evolved over time. In addition, we discovered the popularity and difficulties of pandas topics based on topic views, accepted answers, and score. We display an overview of our methodology in Figure \ref{proposed_methodology}.
\subsection{Data Collection}
We performed our analysis on the dataset that we collected from the Stack Overflow (SO) dump relating to pandas library. We selected SO as it is a site that helps programmers receive facts quicker in the form of a QA site. Since 2008, Stack Overflow consisted over 18 million questions, 11 million users and 51 million different visitors monthly \cite{ref_url2}. We downloaded the SO dump of July 2022 which is the newest during our analysis. We collected a total of 236711 titles of the post containing user-defined tags related to pandas library from the time period July 2008 to February 2022. The titles that were selected had around 144202 accepted answers related to it which is around 61\% of the total titles. Each tuple in our dataset contains the following information: 1. Title of the post 2. Average Score (which is the difference between upvote and downvote of the posts) 3. Average View 4. Accepted Answer Percentage.

\subsection{Topic Modelling using  Latent Dirichlet Allocation} 
Topic modelling can be defined as the function of recognizing topics that best depicts a collection of documents~\cite{topicmodeling}. It is an unsupervised technique of identifying the topics by finding out the nature of the data, like clustering algorithms, which partition the data into different sections. Latent Dirichlet Allocation (LDA) \cite{ref_10} is the most widely used algorithm for Topic Modelling that generates topics based on word frequency in a document. We have used this technique to identify different topics that are present in our dataset.  It generates a topic per document model and words per topic model, designed as Dirichlet distribution. The first step of this algorithm is to model each document as a multinomial distribution of topics and each topic as a multinomial distribution of words. Then it picks the right collection of data by assuming that every chunk of text that is fed into it will hold words that are connected to each other. It also deduces that documents are generated from a combination of topics and topics then produces words based on their probability distribution.

\subsection{Topic Naming}
LDA topic modeling provides topic distributions as the coherence score for each sample. As a result, each sample can be on any of the topics. As such, we select a topic for each sample based on the highest coherence value of the topics following O’callagha et al. \cite{coherance}. Thus, each sample has a topic for which the sample has the maximum coherence score among all topics. Now, each topic has a set of samples. However, these topics have no definitive label as the LDA model only provides scores. Therefore, we label each topic manually by analyzing all the samples. To label these topics, the authors first virtually meet and analyze all the samples for each topic. Then we come up with their possible topic domains and share them with each other. We continue this until all authors unanimously agreed on the topic labeling. For example, a developer asks \textit{`Matplotlib Axes legend shows only one label in barh'} in SO. As this post is related to a Python library that is used for data visualization, we label this post `Data Visualization'. We finally group all of the labeled topics into larger groups by consulting with one another.For example, we find two topics labeled as `Optimization' and `Parallel Processing'. As these topics are related to optimization programming, we group them into a single category titled `Optimizing'.
\vspace{-1.5cm}
\begin{figure}[!ht]
\centering
\hspace*{-3.22cm}
\includegraphics[scale = 0.65]{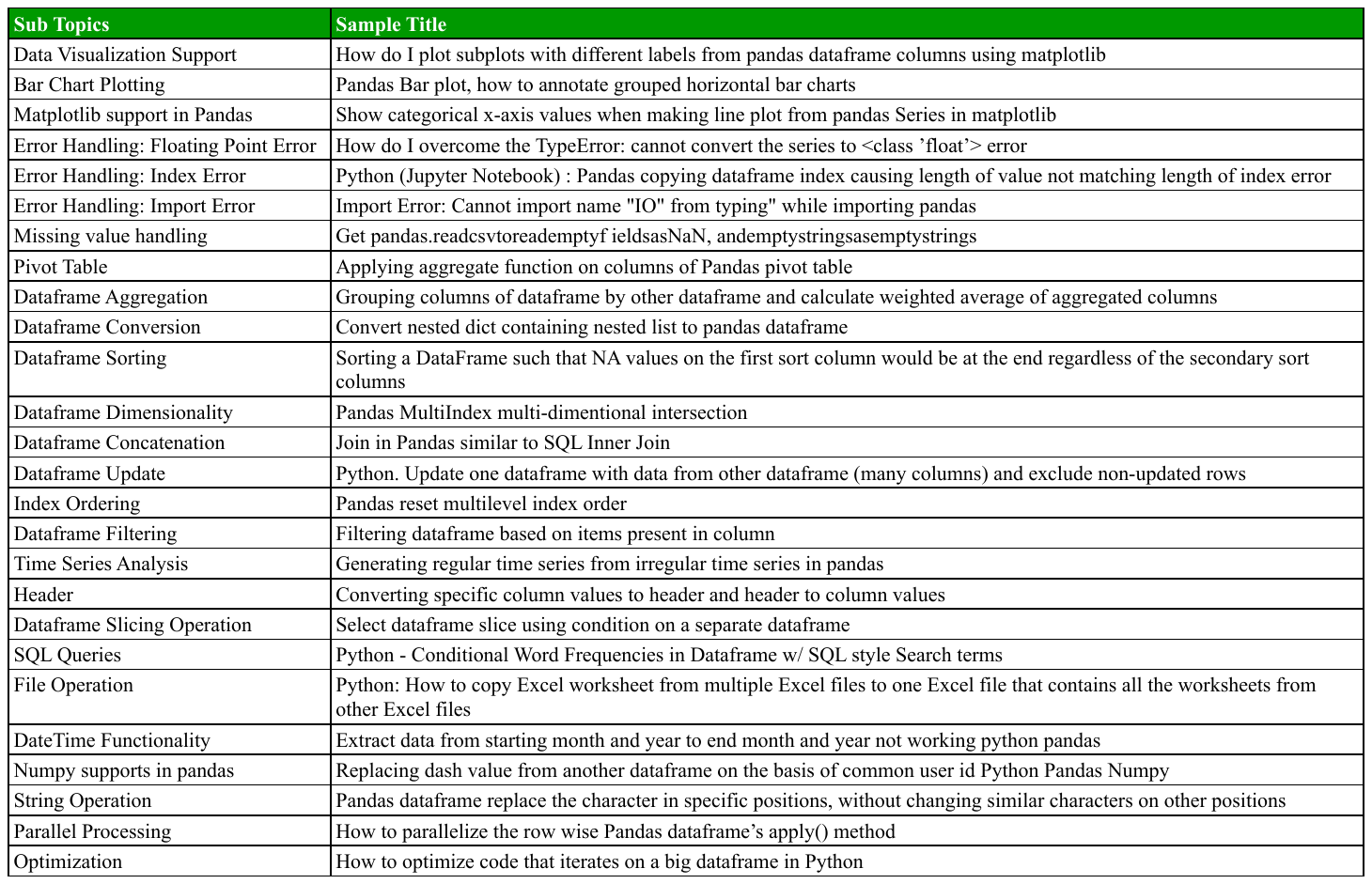}
\hspace*{-3.22mm}%
\vspace{-3cm}
\caption{Sample post title for different categories}
\label{tab:title}
\end{figure}


\begin{figure}[!ht]
\centering
\hspace*{-1.45cm}
\includegraphics[width = 140mm]{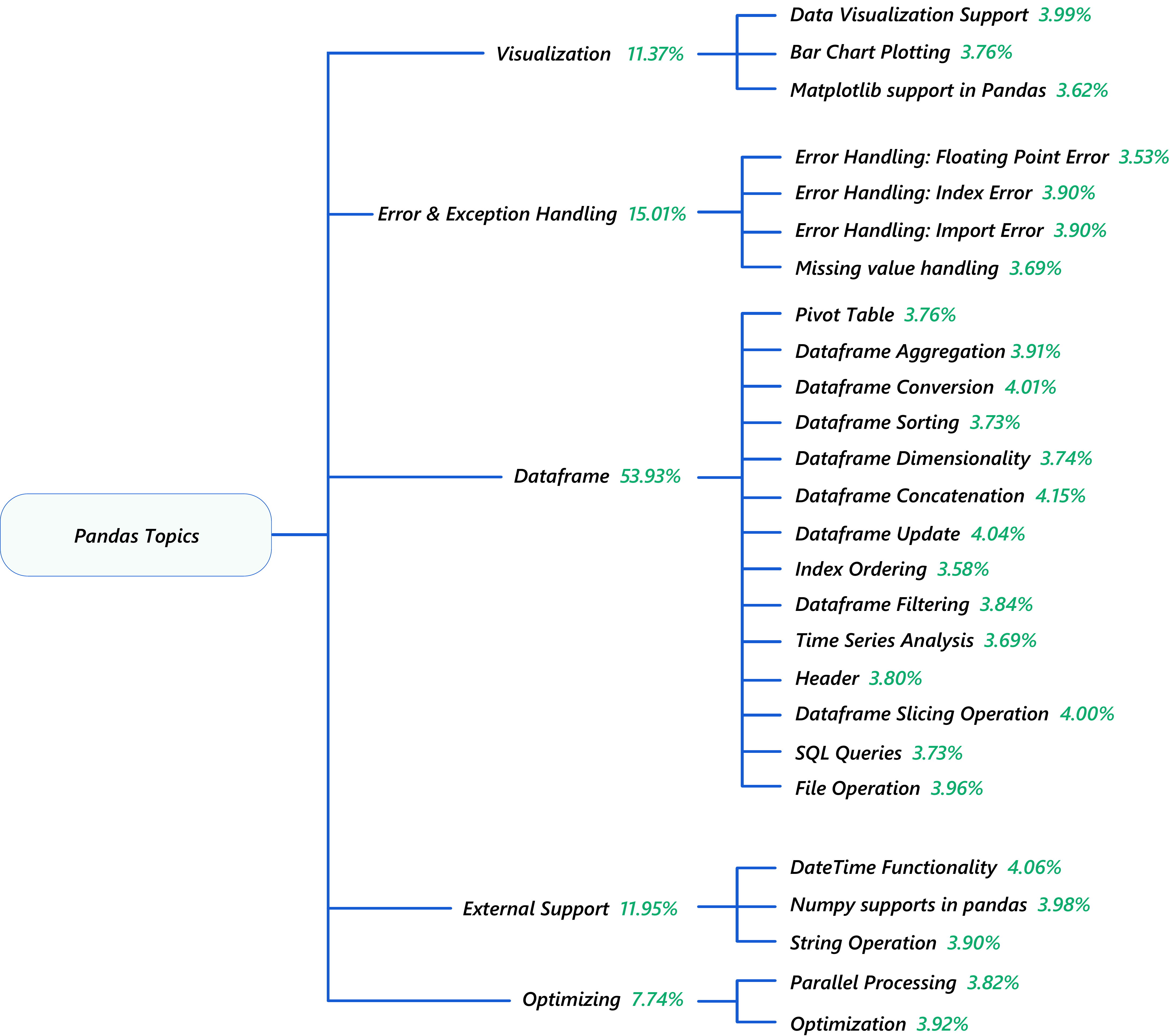}
\hspace*{-1.45cm}%
\caption{Topic Distribution} \label{fig1}

\end{figure}

\subsection{Empirical Analysis}

\subsubsection{Trend Chart}
We used trend chart to illustrate the trend of our topics over time. It depicts the number posts that were made from July 2011 to February 2022 in Stack Overflow on the broad categories of the topics. We calculated number of posts created in each six months for each topics. We constructed line chart using those calculated values. By analysing the trend chart we can observe different characteristics of pandas topics. We can see at what time what specific pandas topic was popular. In short we can find the most trendy topics over different time periods. Also by inspecting the trend chart we can notice whether stack overflow changes with the real world events. We can directly correlate the change in curve of the line graph of the trend chart with change in the real world.

\subsubsection{Topic Difficulties}
Finding difficulties of each topic may help us to detect strenuous topics. By using the information about what topics are difficult, reason behind the difficulties can be investigated. Then by finding out the catalyst of laboriousness, developers can work on and can renovate the library in an easier way. This is why we preferred to measure topic difficulties.  
We explored the difficulties of the topics in our dataset. We analyzed the Score and Accepted Answer Rate of the titles in our dataset to find the difficulty of the titles in our dataset. The titles that have the lowest Score and the lowest Accepted Answer Rate is considered to be the most difficult topic in our dataset. We also further discussed the measures that can be taken to make the difficult topics more familiar to the programmers and developers.

\begin{equation}\label{myeq1}
Avg.~View~for~Topic_i = \frac{{Total~\#~of~views~for~all~posts~in~Topic_i}}{Total~Posts} 
\end{equation}
\begin{equation}\label{myeq2}
Score~for~Topic_i = \frac{Total~Score~of~all~posts~in~Topic_i}{Total~Posts} 
\end{equation}

\begin{equation}\label{myeq3}
Accepted~Answer~Rate~for~Topic_i= \frac{\#~of~Accepted~Answer~in~Topic_i * 100}{Total~posts~in~Topic_i} 
\end{equation}
\subsubsection{Topic Popularity}
As pandas library is expanding day by day, some of the topics are being obsoleted while others are being used more frequently. These used topics or popular topics can give us insightful information, such as the community's demand for pandas topics. This is critical information for newcomers because they must begin with topics that are frequent, easy, and demanding. This helps them keep up to date with the community. Moreover, topic popularity is also useful for the research community to track user views for any new topics. Thus, we studied popularity. To study this, we followed the view count of each post, as there is no concrete formula to calculate popularity. Thus, we calculated Avg. View for each topic using Equation 1. By finding the most popular topics, our model can encourage developers and programmers to focus more on the popular topics or start with the relatively easier topics.

\section{Result and Discussion}
We have divided the Result and Discussion section into 4 parts: Pandas Topic, Evolution of Pandas Topics, Trend Chart Interpretation and Analysing Topic Popularity and Difficulty. In section \ref{sssec:pt}, closer observation on each topic is discussed. About the growth of different pandas topics are discussed in section \ref{sssec:ev} and in section \ref{sssec:tci}, we interpreted the trend of the topics over time with the help of trend chart. Lastly, in section \ref{sssec:tpd} we have analysed topic popularity and difficulty.   

\subsection{Pandas Topic} \label{sssec:pt}
From Fig \ref{fig1} we can observe that the pandas titles were divided into 26 topics and these 26 topics were further categorized into 5 broader topics. Example titles of all the 26 topics have been shown in Table \ref{tab:title}. Our observation on the 26 topics are described below:

\begin{itemize}
\item[o]\textit{\textbf{Data Visualization Support: }}
We found that there exists enough discussion in SO related to data visualization supports, such as subplotting problems, labelling problems, etc. So all the titles that fall under this category were put in this topic. The titles under this topic make 3.93\% of all the titles that we collected.  

\item[o]\textit{\textbf{Bar Chart Plotting: }}
We have observed that 3.76\% of all the titles that we have collected from SO were made on Bar Chart Plotting. Titles containing discussions relating to combining stacks to a Bar Chart, annotating barchart, coloring grouped barchat, etc. were put under this topic.
\item[o]\textit{\textbf{Matplotlib support in Pandas: }}
A lot of discussion relating to Matplotlib support in SO have been made, for example, discussion relating to grid, problems relating to drawing multiple graphs etc. These titles make up 3.62\% of all the titles that we collected. 
\item[o]\textit{\textbf{Error Handling: Floating Point Error: }}
Our SO dataset contained a lot of titles containing Error Handling of Floating Point Error specifically conversion error and retrieving error of floating point values. Posts containing these queries are put under this topic. 
\item[o]\textit{\textbf{Error Handling: Index Error: }}
SO posts containing index error while using pandas library, such as length not matching index, were put in this category.
\item[o]\textit{\textbf{Error Handling: Import Error: }}
SO posts in our dataset containing queries relating to importing pandas library, csv files in pandas, data reader and many other import errors are put in this category.
\item[o]\textit{\textbf{Missing value handling: }} Missing value handling is very important specially for machine learning models. Under this topic, users generally asked for support to fill missing values or drop rows which contains some missing values or NAN value. 
\item[o]\textit{\textbf{Pivot Table: }}
This topic contains posts about aggregating and grouping data in the pivot table.
\item[o]\textit{\textbf{Dataframe Aggregation: }} SO posts under this topic contained queries about executing operations on multiple columns in the dataframe to create new columns in the dataframe.
\item[o]\textit{\textbf{Dataframe Conversion: }}JSON file to dataframe or dictionary to dataframe related queries are can be found under this topic.
\item[o]\textit{\textbf{Dataframe Sorting: }} Sorting dataframe according to certain column, multilevel sorting related consultations are the most common title under this category. 
\item[o]\textit{\textbf{Dataframe Dimensionality: }}Dimentionality related discussions are assigned in this topic.  
\item[o]\textit{\textbf{Dataframe Concatenation: }} User queries related to concatenating two different dataframe with the help of a common column is sorted under this topic. 
\item[o]\textit{\textbf{Dataframe Update: }}Discussions related to updating dataframe, dropping row or column can be found under this topic. 
\item[o]\textit{\textbf{Index Ordering: }}Ordering index properly, multilevel index ordering, index mismatch related discussions can be found most in this category. 
\item[o]\textit{\textbf{Dataframe Filtering: }}Posts related to filtering data according to some criteria are assigned here.
\item[o]\textit{\textbf{Time Series Analysis: }}Preparing dataframe to make it suitable for time series analysis was the main focus of this topic.
\item[o]\textit{\textbf{Header: }}Modifying pandas dataframe header, setting column name, extracting column name these sort of discussions are assigned in this topic. 
\item[o]\textit{\textbf{Dataframe Slicing Operation: }}Discussions related to dataframe slicing are sorted in this topic. This topic contains 4\% of all the posts. 
\item[o]\textit{\textbf{SQL Queries: }}How operations that are analogous to SQL queries can be applied to a pandas dataframe is the key theme of this topic.
\item[o]\textit{\textbf{File Operation: }}Data of pandas dataframe can be saved in different file format like excel. This sort of posts are associated with this topic.
\item[o]\textit{\textbf{DateTime Functionality: }}Formatting date and time from raw string, filtering data that exist within a given time interval are the keywords of this topic.
\item[o]\textit{\textbf{Numpy supports in pandas: }}Numpy is a well known library in python for mathematical operations. A lot of users tried to apply different operations that can be done with numpy on pandas dataframe. This kind of posts lie in this category. This topic contains 3.98\% of all the posts. 
\item[o]\textit{\textbf{String Operation: }}Posts related to modifying string of a particular column are assigned in this category.  
\item[o]\textit{\textbf{Parallel Processing: }}In this category the post of users, who were seeking help to take the advantage of processing data parallelly to make the best use of time, are sorted. 
\item[o]\textit{\textbf{Optimization: }}Same task can be performed by writing different code. In this category, users posted in SO to find optimized way to solve certain problems.

\end{itemize}

We further categorized this topics into five broad categories, i.e., `Visualization', `Error \& Exception Handling', `Dataframe', `External Support', and `Optimizing'. We found that most of the topics are under `Dataframe' category which covers 53.92\% posts. This categories includes all discussion related to `Dataframe Aggregation', `Pivot Table', `Dataframe Conversion', `Dataframe Sorting', `Dataframe Update', `Header', `ataframe Dimensionality', `Dataframe Concatenation', `Time Series Analysis', `Index Ordering', `Dataframe Filtering', `Dataframe Slicing Operation', `SQL Queries', and `File Operations'. The second largest category is `Error \& Exception Handling'. It covers 4 topics and 18.01\% posts. Among the other three categories, `Visualization' and `External Supports' covers 3 topics and 2 topics are covered by `Optimizing' category. We found that developers actively discuss `External Support' while using pandas library. Moreover, we observed that those discussion includes compatibility of pandas library with external libraries, feasibility of using external libraries, performance of those libraries, etc. This indicates that the community should focus on these external libraries to facilitate the usages in pandas library. This requirements demands further investigation on `External Support' in pandas which we left as our future work.

\begin{figure}[!h]
\center
\hspace*{-1.75cm}
\includegraphics[width=\textwidth, height=85mm]{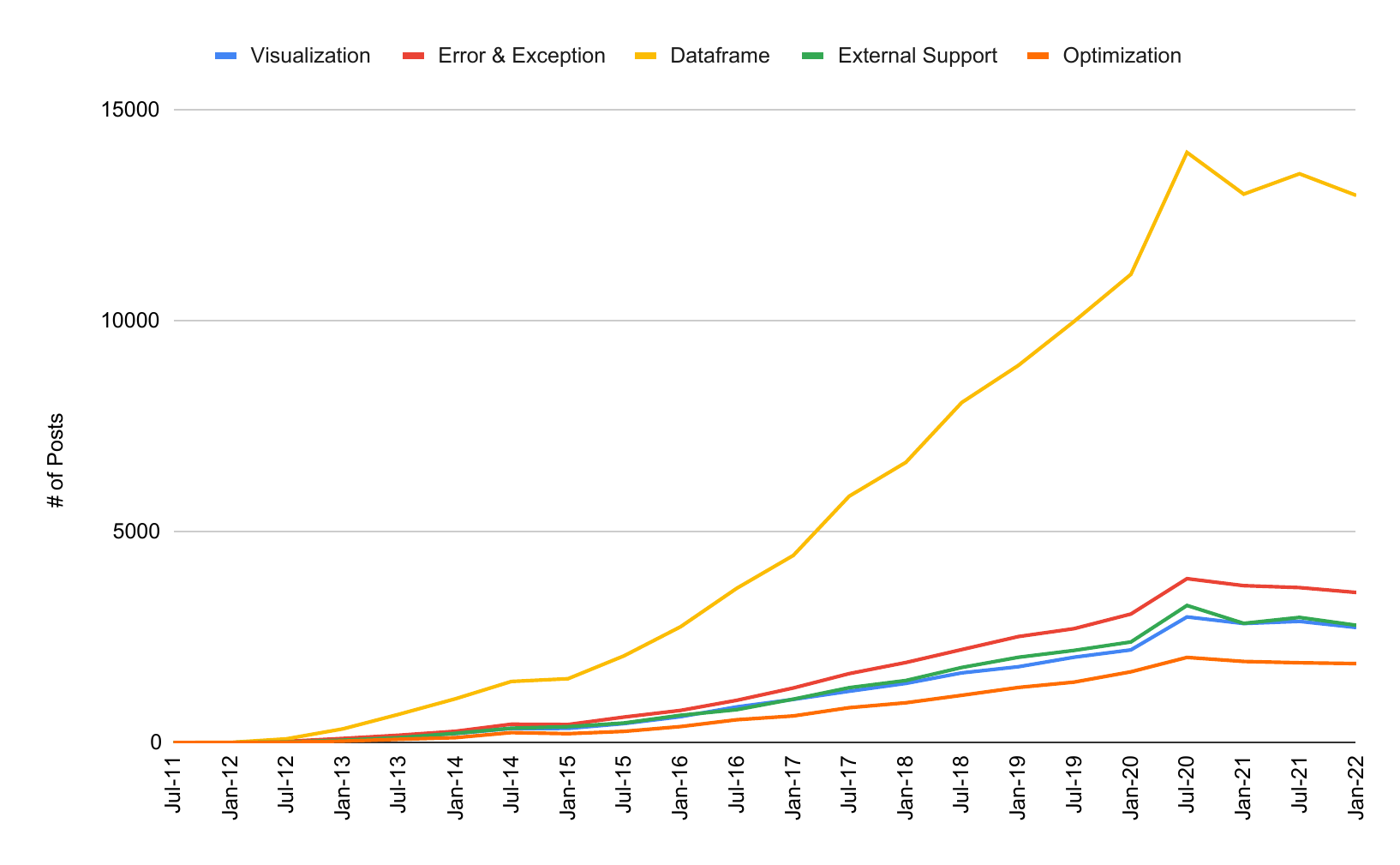}
\hspace*{-1.75cm}%
\caption{Trend Chart} \label{trend_fig}
\end{figure}
\vspace{-1cm}
\subsection{Evolution of Pandas Topics}\label{sssec:ev}
From the trend chart in figure \ref{trend_fig}, it can be observed that the number of posts under different topics gradually increased over time. This behaviour is expected as the community is growing day by day. Dataframe category dominates the others by a large margin, and it continues to grow every year. The least discussed topic is `Optimization'. The distribution of `External Support' category and `Visualization' category is nearly analogous over the given time frame. We also find the second ranked discussion topic is `Error \& Exceptions' which is also piling up as time passes by.

\subsection{Trend Chart Interpretation }\label{sssec:tci}
A closer look at the Figure \ref{trend_fig} notifies us that from January 2017 to July 2020, there is a massive upswing of pandas discussions. This indicates that pandas developers are adopting pandas library for their development work more than ever before. However, it can be noticed that after July 2020, the number of posts for all the categories tends to decline by a very small margin. We investigated this fall of pandas discussions. We find that a massive pandemic outbreak took place in early 2020, which shook the world \cite{covid}. This event forced the world to shut down for a while \cite{covid}. As such, the activity of the developer community reduced by this life threatening virus. As a result, pandas discussions were also hindered and had a downward slope. As a follow up to this event, we find an upward swing in January 2021, when the world get normal for a while \cite{covid}. However, this didn't last long as the virus continued to breakout again during 2021 \cite{covid}. This resulted in another lockdown, and developer activities were hampered as usual. As an immediate effect, we find another drop in pandas discussion in January 2022.
\vspace{-1.9 cm}
\begin{figure}[!ht]
\centering
\hspace*{-3.6cm}
\includegraphics[scale = 0.7]{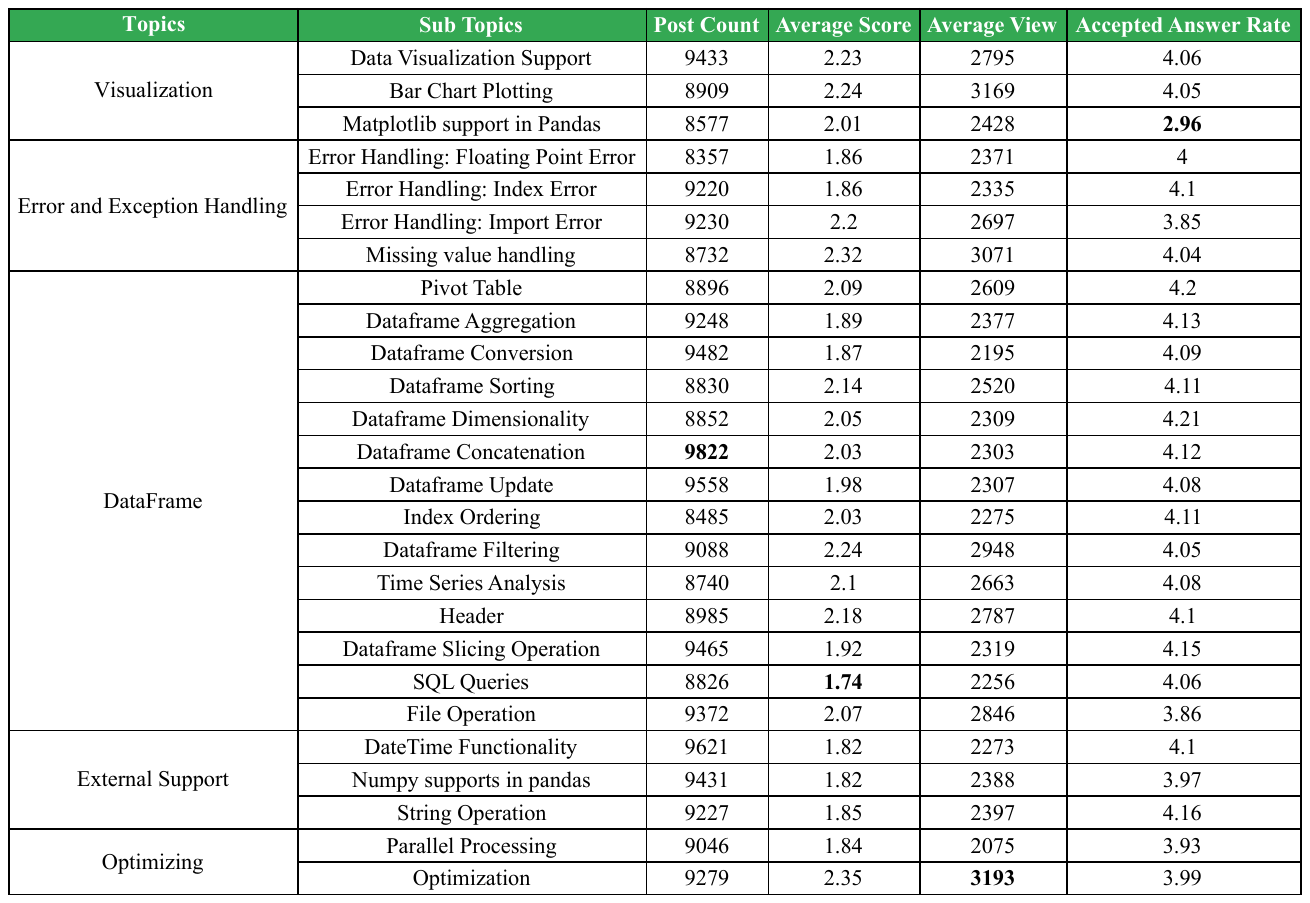}
\hspace*{-3.6mm}%
\vspace{-3cm}
\caption{Topic Hierarchy Statistics}
\label{tab:stat}
\end{figure}
\vspace{-1cm}
\subsection{Analysing Topic Popularity and Difficulty }\label{sssec:tpd}

From Table \ref{tab:stat} we can see that the most popular topic is Optimization as it has the most views. This means the programmer community is more concerned about how they can optimize their code while using pandas library. So, this shows that the developers need to emphasize more on the functionalities of optimization of pandas library to create a better coding experience for the programmers using this library. The researchers are also interested to make their research more optimized so that they can make their model more efficient, so the developers can collaborate with the researchers to make the optimization part of pandas library more handy for the research users. We can also see the that second most popular topic is Bar Chart Plotting as it has the second most views. The least most popular topic is SQL queries as it has the least amount of views. As SQL queries is a topic of database, it is not that important in the domain of pandas library. So developers show less interest in this specific topic.

From Table \ref{tab:stat}, we can see that the topic containing the highest Average Score of 2.35 is Optimization and the lowest Average Score of 1.74 is SQL Queries. The topic Matplotlib support in Pandas contains the lowest Accepted Answer Rate of 2.96 and Dataframe Slicing Operation contains the highest Accepted Answer Rate of 4.15. We can also observe from Table \ref{tab:stat}  that the most difficult topics in pandas library are SQL queries as it has the lowest Score  and Matplotlib support in Pandas as it has the lowest Accepted Answer Rate. This means the general programmers struggle most in these topics. So the experts in pandas library can analyze these topics more to make the programmers use these topics more effectively.

\section{Implication of Studies}
Several important expositions can be observed from our work. 
\begin{itemize}
\item[o]\textit{\textbf{Learner: }}The enthusiastic learners who are trying to digest pandas library, can gain some idea about the popularity of different topics associated with pandas. This will paint a pathway for the learners to grasp pandas library in a more structured manner. To speed up their learning process they can learn relatively easier topics first. They can also prioritize their learning process according to the popularity of the topic if they want. 
\item[o] \textit{\textbf{Research Community: }}Researchers, developers, marketers can analyze and decide what topics questions are frequently asked, and on what topics questions are frequently answered. Developers can see also which topics are popular for marketing purpose so that they can add new handy features on that. Rate of accepted answers can also be seen from our work by means of which the developers can work on the topics that has less accepted answers.
\item[o]\textit{\textbf{Educators: }}Instructors community of pandas can understand which topics need to be more focused based on the difficulty level of different topics. They can analyze the more difficult topics and make materials on it to better educate the learners on those specific topics. They can also contribute to the queries that has lower number of accepted answer.
\item[o]\textit{\textbf{Data science community }}Those who are involved in data analysis can contribute more to the community by finding difficult problem and solve them in a easier convenient way. The community can also observe the trends of the evolving topics related to pandas.   

\end{itemize}
\section{Threats to validity}
There are certain factors that can impede the righteousness of our work. If any complementary library of pandas arises in future, our model will not be as impactful as it stands today. Then the study can be shifted to the another library analogous to pandas. Besides, we have considered the most upvoted posts while structuring our model. If this system changed, our model may not be valid anymore. Because with the change of scoring system our model could not relate to the new scoring system. Moreover, if some portion of data that we have used in our model is erased by Stack Overflow, our model will not be irrefutable. In addition, while modeling different topics, we assigned each title to a certain topic by manual observation. But there were cases where a single title might be categorized as multiple topics that avert our models purity. The density of this cases  

\section{Future Work}
Our work could be expanded by investigating the popular and difficult topics further. We plan to break down the difficult topics to a greater extent so that we can understand the tendency of the general programmer, whether the reason for the those topics to be difficult is actually it's difficulty or reluctance of the general programmers to learn those topics. 

\section{Conclusion}
In this study, we explored pandas library associated argument on Stack Overflow (SO) and employed topic modeling to institute different topics. We have added several findings about it. Pandas learners discuss a variety of topics in SO related to Visualization, Error and Exception Handling, DataFrame, External Support and Optimization.  All these topics are exploding swiftly. Among them Dataframe is the most discussed topic because of its huge usage. Impactful factors like topic difficulties, topic usage, post count of different can be observed from our work which can act as the catalyst for the different pandas contributors to improve techniques, tools, and methods related to pandas.

%
%
%
%

\end{document}